\documentclass[12pt]{article}
\usepackage{epsfig}

\parskip=\medskipamount 
plus.3em minus.5em \abovedisplayshortskip=.5em plus.2em minus.4em
\renewcommand{\thesection}{\arabic{section}}

\catcode`@=11

\@addtoreset{equation}{section} \@addtoreset{equation}{subsection}
\def\theequation{\ifnum\value{section}=0 \arabic{equation}\ignorespaces
\else \ifnum\value{section}=-1 A.\arabic{equation}\ignorespaces
\else \ifnum\value{subsection}=0
\thesection.\arabic{equation}\ignorespaces \else
\thesection.\arabic{subsection}.\arabic{equation}\ignorespaces
                             \fi
                        \fi
                   \fi}

{\catcode`\'=\active \def'{{}^\bgroup\prim@s}}

\catcode`@=12

\newcommand{\bq}{\begin{equation}}
\newcommand{\be}{\begin{equation}}
\newcommand{\fq}{\end{equation}}
\newcommand{\ee}{\end{equation}}
\newcommand{\bqr}{\begin{eqnarray}}
\newcommand{\beqs}{\begin{eqnarray}}
\newcommand{\fqr}{\end{eqnarray}}
\newcommand{\eeqs}{\end{eqnarray}}

\newcommand{\rf}[1]{(\ref{#1})}

\def\bop#1{\setbox0=\hbox{$#1M$}\mkern1.5mu
    \vbox{\hrule height0pt depth.04\ht0
    \hbox{\vrule width.04\ht0 height.9\ht0 \kern.9\ht0
    \vrule width.04\ht0}\hrule height.04\ht0}\mkern1.5mu}

\begin{document}
\thispagestyle{empty}

\begin{flushright}
\begin{tabular}{l}
\end{tabular}
\end{flushright}

\vskip .6in
\begin{center}

{\bf Simplifications of the Tensors in Quantum Scattering, Part I}

\vskip .6in

{\bf Gordon Chalmers}
\\[5mm]

{e-mail: gordon@quartz.shango.com}

\vskip .5in minus .2in

{\bf Abstract} 

\end{center}  

In this comment it is pointed out that the perturbative dynamics of 
general massive field theories can be mapped to delimited sums of 
determined integral functions.  The limits on the sums are the remaining 
obstacle to finding the general $g$-loop, or derivative expanded, 
form of the scattering.  

\vfill\break

In previous works the classical scattering of scalar and gauge 
field theories has been formulated with the use of partitions 
of numbers \cite{ChalmersA},\cite{ChalmersB}.  The use of the 
classical scattering has been used in the formulation of the 
perturbative quantum dynamics \cite{ChalmersC},\cite{ChalmersD}.  
There are simplifications in the quantum scattering with the 
use of the determined integral functions, presented in 
\cite{ChalmersC}.  The content of this letter is to show that 
the perturbative dynamics of a general massive theory can 
be reduced to the, as yet delimited, sums of numbers 
parameterizing the products of these integral functions.  

In massive $\phi^3$ theory, with coupling $g$, the scattering takes on 
the form \cite{Chalmers3}, 
\bqr 
g^L m^q ~\prod s_{ij}^{n_{ij}} \sum \prod_i B(n_i,m_i,w_i) \ , 
\label{expansionscalar}
\fqr 
with the parameters $n_i$, $m_i$ and $w_i$.  There are conditions on 
the sum which depend on the loop order $L$ and the kinematic configuration of 
$n_{ij}$.  The $B$ functions are integral moments defined and calculated in 
\cite{Chalmers1}. 

The simplest example is symmetry breaking in QED, 
\bqr 
{\cal L} = -{1\over 4} F^2 + {1\over 2} \nabla \phi^\dagger \nabla \phi + 
   {\bar\psi}^a \gamma^\mu\nabla_\mu\psi_a \ , 
\fqr 
with $\nabla_\mu=\partial_\mu+g A_\mu$; the masses of the fermion 
and scalar are left non-zero.  The vacuum value of the scalar is 
chosen as, 

\bqr 
\nabla\phi^\dagger\nabla\phi=\partial\phi^\dagger\partial\phi + 
2g A(v+\phi)^\star\partial\phi+c.c. + g \vert v+\phi\vert^2 A^2 \ , 
\fqr 
which gives a mass to the gauge field. 

The non-abelian guage theory is defined by, 
\bqr 
{\cal L} = -{1\over 4} F^b F_b + {1\over 2} \nabla\phi^a\nabla\phi_a  
\fqr 
with the covariant derivatives, 

\bqr 
\nabla=\partial+g A \qquad  \nabla_\mu\phi^a = \partial_\mu \phi_a 
+ g  [\phi,A_\mu]_a  \ . 
\fqr 
The symmetry breaking could be achieved as in the standard $SU(N)\rightarrow 
U(1)$, with additional interactions to give masses to the abelian fields. 

One example of symmetry breaking is as follows,

\bqr 
\phi_a \rightarrow \phi_a+v_a
\fqr 
with, 
\bqr 
\partial\phi^a\partial\phi_a + g^2 f^{cab}f_c^{de} 
 (v_a+\phi_a)(v_d+\phi_d) A_b A_e    
\fqr 
\bqr 
+ g 2 {\rm Tr}T_c[T_a,T_b] ~(v_b+\phi_b)A_c \partial^\mu \phi^a  \ . 
\fqr 
The mass term is 

\bqr 
(v_a+\phi_a)(v^a+\phi^a) A_b A^b - [(v_a+\phi_a)A^a]^2 
\fqr 
The vacuum expectation values are chosen uniformly $v_a=v \eta_a$ 
($\eta^2=1$), so that 
\bqr 
v^2 A^2 - v^2 \eta\cdot A^2 = v^2 [A^2 - (\sum \eta_a A_a)^2] 
\fqr 
\bqr 
=  v^2 \sum (1-\eta_a^2)A_a^2 - 2 v^2 \sum \eta_a \eta_b A_a A_b  \ , 
\fqr 
which can be used to give mass to all of the the gauge fields.  The 
gauge fixing term could be written in axial gauge as 
$v^2(\eta\cdot A)^2$, and makes the uniform breaking manifest. 

In three dimensions, a gauge invariant mass could be generated with the 
topological term 
\bqr 
m^2 F\tilde F \ , 
\fqr 
without any use of the scalar fields.  

The gauge scattering, with $n$ external gauge field states, then takes on 
the form, 
\bqr 
\prod_{i=1}^m \varepsilon_i\cdot \rho_j ~ g^q~
   {m_A^{p_1} m_\phi^{p_2}} ~ \prod s_{ij}^{n_{ij}} 
 \sum \prod_i B(n_i,m_i,w_i) \ , 
\label{expansionnonabelian}
\fqr 
with the possible tensors 
\bqr 
\rho_j = \epsilon_j \quad{\rm or}\quad k_j \ .  
\label{helicityconf}
\fqr 
The complete quantum gauge scattering then is produced by the appropriate 
summations on the indices in the sum in \rf{expansionnonabelian}.  This 
expansion 
depends on the helicity configuration in \rf{helicityconf}, that is 
choosing the contractions of momenta or helicities as $\rho_i$, with $m$ 
ranging from $n/2$ to $n$ (all polarizations or all momenta as $\rho_i$).

In the example of a three dimensional gauge field with the mass term, the 
form is simpler, 
\bqr 
\prod_{i=1}^m \varepsilon_i\cdot \rho_j ~g^L 
   m^p \prod s_{ij}^{n_{ij}} \sum \prod_i B(n_i,m_i,w_i) \ , 
\label{expansionqed}
\fqr 
with the indices on the sum reflecting the dynamics of the three dimensional 
massive gauge theory.  The summations on the $B$ functions depend on the 
helicity configuration and the powers of the coupling.

In any theory with the mass generation for the gauge fields, and with 
non-zero values for the remaining particle content, there is an expansion 
of the form \rf{expansionscalar}, \rf{expansionnonabelian}, or 
\rf{expansionqed}, with the appropriate mass values used in their forms.  
The integral functions $B(n_i,m_i,w_i)$ are determined.  The completion 
of the quantum scattering is determined by the appropriate summations of 
the indices.  The required non-analyticity due to unitarity, such as 
$\ln(s-m^2)$ is found by resumming the polynomial terms.

These summations on the indices are potentially determined through a 
simple context, such as the use of some analog of the Hopf algebra associated 
with the tree diagrams.  The determination of the sum limits in the scattering 
determines the perturbative quantum scattering, with possible redundancy 
in the sums, and in a general massive theory with various particle content.

\vfill\break

\vskip .2in

\vfill\break

\vfill\break

\vskip .6in
\begin{center}

{\bf Tensor Simplifications, Part II: Amplitude Parameterizations}

\vskip .6in

{\bf Gordon Chalmers}
\\[5mm]

{e-mail: gordon@quartz.shango.com}

\vskip .5in minus .2in

{\bf Abstract}

\end{center}

The amplitudes in gauge theory, gravity, and gauge theories with 
matter are parameterized.  The coefficients of the parameterized 
amplitudes are required in order to specify the complete coupling 
form, and in determining the couplings within the derivative 
expansion.  However, these parameterizations can be used in a 
similar manner as a Fourier transform in a fit to 
experimental data over large energy ranges.

\setcounter{page}{2}
\newpage
\setcounter{footnote}{1}

The computation of gauge theory amplitudes is typically complicated 
due to both the loop integrals and the tensor computations.  Many tools 
have been developed to circumvent these difficulties.  The computation 
of gauge theory amplitudes has been extensively studied for many years. 
The parameterization of the amplitudes is useful in their computation 
within the derivative expansion \cite{Chalmers0}-\cite{Chalmers16}. 
The general form of the gauge theory amplitudes has uses in particle 
phenomenology.  

The parameterization of the amplitudes has never been elucidated in a 
clear context in the literature.  
The general form of the the pure gauge theory amplitude is 

\bqr 
\prod \varepsilon_{\sigma(i)} \cdot \varepsilon_{\tilde\sigma(j)} 
\prod \varepsilon_{\rho(i)} \cdot k_{\tilde\rho(j)}   
\fqr 
\bqr 
{\prod t_{a(i)}^{{\tilde a}(i)} \over \prod t_{b(i)}^{{\tilde b}(i)}}  
\prod \ln^{d_1(i)} \ldots \ln^{d_m(i)} t_{c(i)}^{{\tilde c}(i)}  \ .  
\label{expandedform} 
\fqr 
The sets $\sigma$, $\tilde\sigma$, $\rho$, $\tilde\rho$, 
$a$, $\tilde a$, $b$, $\tilde b$, $c$, 
$\tilde c$, $d_l$ are required in order to specify the general form 
of the amplitude.  In the case of massive theories there are factors 
of the mass parameters.  The inclusion of the fermions requires the 
factors 

\bqr 
\prod \psi_a^\alpha \psi_{b\alpha} \prod {\bar\psi}_a^{\dot\alpha} 
 {\bar\psi}_{b\dot\alpha} 
\prod \psi_i^\alpha k_{\alpha\kappa(i)} 
 {\bar\psi}^{\dot\alpha,i} k_{\dot\alpha\dot\kappa(i)} 
\prod \psi_i^\alpha {\bar\psi}_j^{\dot\alpha}
 \varepsilon_{\alpha\dot\alpha}(\nu(i)) \ , 
\fqr 
which extends from the external line factors.  

There is a further group theory tensor, 

\bqr 
\prod_j {\rm Tr} \prod_i T_{\gamma_j(i)} \ , 
\fqr  
with as many traces there are loops in the amplitude calculation.  
The pure gravity amplitude, also in perturbation theory, requires 
the same form, except there are are no traces; there is a doubling 
of the $\varepsilon$'s into holomorphic and non-holomorphic parts, 
$\varepsilon_i$ and ${\bar\varepsilon}_i$.  

There is a numerical factor multiplying the tensor form, and is 
proportional to the gauge coupling, $g^{n-2+2L}$.  This numerical 
coefficient is required to specify the complete amplitude, and 
at each loop order.  

The numerical factor 

\bqr 
f(g,M)=\sum b_L(M) g^{n-2+2L} \ ,  
\label{coupling}
\fqr 
with $M$ representing the set of values $\sigma$, $\rho$, $a$, $\tilde a$, 
$b$, $\tilde b$, $c$, and $\tilde c$, $d_l$.  These coefficients can be 
iteratively determined in the gauge theory derivative expansion.  There 
is a restriction on the number of logarithms due to the number of loops.  
Possible functional forms of the scattering are, 

\bqr
\rho_M \sum \ldots \sum F_{p,q} \ldots F_{p,q}  \ , 
\fqr 
for the gauge scattering, and termwise for the graviton scattering 
(which has a dimensionful coupling constant), 

\bqr 
\gamma_M \Gamma \ldots \Gamma / \Gamma \ldots \Gamma \ . 
\fqr 
These forms are line with lower dimensional field theories.  

However, despite the determination of the amplitudes in a straightforward 
calculational scheme, their form may be used to model partonic cross-sections; 
the linear combination of the terms in \rf{expandedform} can be used to 
define cross-sections in the usual manner.   In other words, a large set 
of these terms in \rf{expandedform} can be used 
to model a greater range in energies in particle amplitudes used in,  
for example, QCD or electroweak cross sections.  Their form, as a collective 
set, can model very large ranges in energies, even without knowing the 
explicit coupling dependence in \rf{coupling}.    

An interesting further question is the inversion of a general linear 
combination of the terms in \rf{expandedform} into an effective Lagrangian 
and its properties. 

\vfill\break

\end{document}